\documentclass[11pt, oneside]{article}   	
\usepackage{geometry}                		
\usepackage{graphicx}				
\usepackage{amssymb}
\usepackage{amsmath}
\usepackage{natbib}
\usepackage{xcolor}

\usepackage[pdftex,bookmarks,colorlinks,breaklinks]{hyperref}
\definecolor{dullmagenta}{rgb}{0.4,0,0.4}   
\definecolor{darkblue}{rgb}{0,0,0.4}
\hypersetup{linkcolor=blue,citecolor=darkblue,filecolor=dullmagenta,urlcolor=darkgreen} 

\newcommand{\lambdabar}{{\mkern+04mu\mathchar'26\mkern-09mu\lambda}}

\def\surconduct{{\sigma^s}}
\def\surresist{{R^s}}
\newcommand*{\excess}[2][]{{#2}_{#1}^s} 
\def\surfacelength{{\ell_0}}
\newcommand{\bulk}[1]{_{\mathrm{#1}}}
\def\OmegaP{\Omega_{p}}
\def\pol{\mu}

\usepackage{times}

\newlength{\extendtextwidth}
\renewcommand{\caption}[2][]{\refstepcounter{figure}%
\vspace*{1ex}\hspace*{8.5ex}%
\parbox{0.8\textwidth}{\small
\textbf{\figurename~\arabic{figure}.} #2}}

\title{\Large\bf No-slip boundary conditions for electron hydrodynamics\\ 
and the thermal Casimir pressure}

\author{Mandy Hannemann$^1$, Gino Wegner${}^{2,3}$ and Carsten Henkel$^1$
\\[1ex]
\small
${}^1$ University of Potsdam, Institute of Physics and Astronomy,
\\[-0.5ex]
\small
Karl-Liebknecht-Str.~24/25,
14476~Potsdam, Germany
\\
\small
${}^2$ Humboldt-Universit\"at zu Berlin, Institut f\"ur Physik,
\\[-0.5ex]
\small
AG~Theoretische Optik \& Photonik, 
12489~Berlin, Germany
\\
\small
${}^3$Institute of Condensed Matter Theory and Optics, Friedrich-Schiller-University Jena,
\\[-0.5ex]
\small
Max-Wien-Platz~1, 07743 Jena, Germany
} 
\date{\small 2021 Apr 20}

\begin{document}
\maketitle

\abstract{%
\centerline{\parbox{0.8\textwidth}{\noindent
We derive modified reflection coefficients for electromagnetic waves in the THz and 
far infrared range. The idea is based on hydrodynamic boundary conditions for metallic
conduction electrons. The temperature-dependent part of the Casimir pressure between metal plates
is evaluated. The results should shed light on the ``thermal anomaly'' where measurements
deviate from the standard fluctuation electrodynamics for conducting metals.
\\[1ex]
}
}}

\normalsize

\section{Introduction}
\label{s:intro}

The Universe is mainly filled with matter and radiation. While the former makes up
a mass of roughly one Hydrogen atom per cubic meter, averaged over cosmological scales,
quantum theory predicts since about one century ago
that the zero-point energy of radiation, one half photon energy
per electromagnetic mode, sums up to an energy per unit volume of
\citep{Adler_1995}
$$
u = \frac{2}{V} \sum_{k} \frac{\hbar c k}{2} \sim \hbar c \, \Lambda^4
$$
where the factor $2$ accounts for two transverse polarizations, and 
$\Lambda$ is a cutoff of the $k$-space available to the modes. 
Choosing this at the Planck length $1/\Lambda = (\hbar G/c^3)^{1/2}$,
one obtains an energy density a factor $\sim 10^{123}$ above the energy equivalent of
the matter content: the ``wrongest formula in physics''. 
It is intriguing to handle this discrepancy with an argument familiar from 
renormalization theory (and also used by \citet{Casimir_1948b}): 
the vacuum energy scales in leading order
with the volume of the system and can be subtracted by comparing two situations with 
the same volume -- but differing in boundary conditions. However, what should set 
boundary conditions for the Universe as a whole and where? -- perhaps the cosmological 
horizon, effectively considering
the Universe as a bubble with Hubble radius $c/H$. Subtracting the energies, a term 
remains that scales with the surface of the bubble. Taking the short-wavelength cutoff
$\Lambda$ for $k$-vectors parallel to the horizon, the vacuum energy density
in our Universe bubble becomes
$$
u \mapsto \frac{ \hbar c \, \Lambda^2 }{ (c/H)^2 } \sim \frac{ H^2 }{ G / c^2 }
$$
The subtraction has removed $\hbar$ from the formula (if the Hubble constant $H$ 
is considered a given parameter), and it has reduced the radiation energy by a factor 
$(H / c \Lambda)^2 \sim 10^{-122}$ -- leading to an estimate
comparable to the observed mass-energy. For other estimates about
the cosmological horizon and its role for the possibly accelerated expansion,
see \citet{Easson_2011}. An analysis of the self-gravitation of the huge vacuum energy
density has been given by \citet{Wang_2017c}.

The above estimate cannot serve more than to illustrate how difficult it is to come to
grips with quantum field fluctuations. This is where laboratory measurements of the Casimir
force come into play. 
The Casimir effect predicts a (generally) attractive force between macroscopic objects,
due to quantum fluctuations of the surrounding fields
\citep{Casimir_1948b, Sernelius_2018book}. Direct measurements of these so-called dispersion forces
are often popularised as improving our
understanding of vacuum fluctuations -- see, however, the viewpoint of \citet{Jaffe_2005} who
recalls the tracing-back to zero-point fluctuations of charges and currents in matter.
A related motivation drives experiments searching for
fundamental corrections to short-range forces that arise
from axion fields or compactified dimensions \citep{Chen_2016b, Klimchitskaya_2021}.
Advances in this field are somehow stalled, however, by the relatively 
down-to-Earth issue of how to characterize precisely the electromagnetic Casimir forces 
between real metallic conductors. Metallic objects have obvious advantages because electric
forces can be avoided which typically mask the weaker dispersion interactions. The problem
arises, however, that the Casimir force under realistic conditions also contains a 
temperature-dependent contribution (sometimes this is attributed to ``real'' rather than
``virtual'' particles) whose relevant frequencies peak in the infrared 
($\hbar \omega \sim k_B T = 25.2\,{\rm meV}$ at room temperature, e.g., Wien's displacement law)
\citep{Torgerson_2004, Bimonte_2009b}.
Its evaluation requires the knowledge of the infrared conductivity of a metal,
but this has been the subject of great discussions, 
the ``Drude vs. plasma'' controversy \citep{Mostepanenko_2015, Henkel_2020a, Reiche_2020a, Klimchitskaya_2020b, Klimchitskaya_2020a}.

The bulk conductivity, however, provides only half
the answer to the response of a conducting object to an external field: its surface 
and its geometry play equally important roles. 
The controversy has taught us that frequencies in the thermal range
do give a significant contribution to the Casimir force, on the one hand. 
On the other hand, from
experience with calculations, the $k$-vectors of the 
relevant field modes are set by the (smallest) distance $d$
between the objects, typically much shorter than the thermal wavelength
$\lambdabar_T = \hbar c / k_B T$ (at room temperature $\sim 7.5\,\mu{\rm m}$).
We are facing a curious combination of $k$ and $\omega$:
contrary to our intuition about infrared frequencies, 
the relevant length scales are below $100\,{\rm nm}$, being determined by $d$.
This calls for a reappraisal of methods
that have been developed over the last century.
Those coming from the context of
infrared spectroscopy do not address the range of parameters $k \gg \omega/c$,
since one is dealing with the response to long wavelengths
$\lambda = 2\pi c / \omega$. More relevant is work on electron energy loss spectroscopy \citep{Verbeeck_2005}
where the fields correspond to the Coulomb potential of a moving charge. 
In that context, however,
the focus has been on rather high frequencies (energies), even the surface plasmon resonance 
(in the visible or UV) being considered a low-energy feature
\citep{Zangwill_1988_chap7}.
The situation is exacerbated by 
experiments addressing,
in the distance range $d \sim 10\ldots 300\,{\rm nm}$,
the Casimir force \citep{Chen_2016b, Bimonte_2016}
and non-contact heat transfer \citep{Kloppstech_2017, Cui_2017a}:
they give results that disagree with the standard theory of 
the fluctuating electromagnetic field \citep{Landau_vol9, Rytov_vol3}. 

We outline in this paper an improved, hydrodynamic approximation for the electromagnetic
response of conduction electrons at a metallic surface. It is shown in particular
that the classical Fresnel formulas based on a local dielectric function apply only in 
a specific range in the $k\omega$-plane, as it happens also with other models including
spatial dispersion \cite{garcia1979introduction, DresselGruenerBook}.
In our context, a kind of conspiration of scales has to be addressed.
To fix a relevant set of parameters, consider 
the commonly used local Drude dielectric function and conductivity
\begin{equation}
\varepsilon\bulk{m}( \omega ) = \varepsilon\bulk{b} - \frac{ \Omega_p^2 }{ \omega (\omega + {\rm i} / \tau) }
\,,\qquad
\sigma(\omega) = \frac{ \sigma_0 }{ 1 - {\rm i} \omega \tau }
\label{eq:Drude-model}
\end{equation}
where $\varepsilon\bulk{b}$ (possibly frequency-dependent, too) describes the response of
bound electrons, the plasma frequency $\Omega_p$ scales with the root of the conduction
electron density, and $\tau$ is the scattering time. The latter can be determined from the
DC conductivity $\sigma_0 = \varepsilon_0 \Omega_p^2 \tau$. Typical parameters for gold
at room temperature
are $\hbar\Omega_p = 9.1\,{\rm eV}$ and $\hbar/\tau = 27\,{\rm meV}$
(wavelength $46\,\mu{\rm m}$, in the far infrared).
The scattering rate separates
the classical Hagen-Rubens regime (low frequencies) from the so-called relaxation regime
$1/ \tau \ll \omega \ll \OmegaP$ (see \citet{Sievers_1978}
and \citet[Appendix~E]{DresselGruenerBook} for more details).
The first coincidence is that the typical thermal frequency is quite close to the Drude 
scattering rate $k_B T / \hbar = 0.94/\tau$ (gold at room temperature).
(For a detailed study of the behaviour of Casimir pressure and entropy at low temperatures,
see \citet{Intravaia_2010c, Reiche_2020a}.)
The second coincidence is one of length scales. 
Recall that an electromagnetic field in the thermal frequency 
band penetrates into a metal in a diffusive way, leading to the characteristic length 
$l\bulk{m} = (\hbar D\bulk{m} / k_B T)^{1/2}$
where $D\bulk{m} = 1/(\mu_0 \sigma_0)$ ($\mu_0$ being the permeability)
is the diffusion constant for magnetic fields \citep{jackson2014klassische}.
The scale $l\bulk{m} \sim 20\,{\rm nm}$ explains the ``thermal anomaly'' of the 
Casimir pressure, namely that 
temperature-dependent corrections appear already at distances $d$ much shorter
than $\lambdabar_{T}$ \citep{Bostrom_2000b, Intravaia_2009a}.
The hydrodynamic model introduces another length in the same range, namely 
the electronic mean free path $\ell$. 
From the Fermi velocity $v_{\rm F}$ (gold: $1.4 \times 10^{6}\,{\rm m/s}$),
we have $\ell = v_{\rm F} \tau = 34\,{\rm nm}$. This scale
appeared already in the anomalous skin effect 
\citep{DresselGruenerBook}
that occurs when $\ell$ is larger than the classical penetration depth 
$(c/\omega)/ \mathop{\rm Im} \sqrt{\varepsilon\bulk{m}}$.
Its impact on the Casimir pressure has been studied using generalisations of the Fresnel
reflection amplitudes \citep{Esquivel_2003, Svetovoy_2006a}, although the modifications
were found to occur only for p-polarized field modes.
This polarization contains an electric field component perpendicular to the surface 
and probes the charge density profile at a metallic surface, whose
characteristic scale is the Fermi wavelength $2\pi / k_{\rm F} = 5.2\,\text{\AA}$
or the Thomas-Fermi screening length $v_F / \Omega_p = 1.0\,\text{\AA}$
\citep{Feibelman_1982, Wegner_2020}.

We focus our hydrodynamic approach on the response to s-polarized fields whose
electric field is parallel to the surface. 
To understand the basic idea, recall
the Maxwell matching conditions for the tangential electric field in vacuum and metal, 
${\bf E}_\Vert({\rm vac}) = {\bf E}_\Vert({\rm m})$.
Using Ohm's law in local form ${\bf j}_\Vert({\rm m}) = \sigma\,{\bf E}_\Vert({\rm m})$,
a nonzero current density right at the metal surface appears. 
This is not consistent with the 
no-slip boundary condition typical for the hydrodynamics of viscous fluids \citep{landau1987fluid}.
For electrons moving parallel to a surface, the no-slip condition takes into account,
on length scales much larger than the Fermi wavelength, the scattering by surface roughness
and by amorphous reconstructions of the sub-surface material. 
Note that this would not apply for atomically clean samples, but rather for metals
kept in ambient conditions, as also suggested by experiments in the THz range \citep{Laman_2007}.
Our hydrodynamic calculations indeed predict that the electronic current density varies significantly 
in the sub-surface region on the scale of the mean free path $\ell$.

A key parameter is the shear viscosity of the electron fluid. We use the observation of 
\citet{Conti_1999}
that the hydrodynamic (Navier-Stokes) equations can be phrased in the language of 
a visco-elastic medium. 
An elastic modulus and the viscosity are related, 
at finite frequencies, to the real and imaginary parts of the same mechanical response.
It turns out that this response is encoded in the longitudinal and transverse
dielectric functions of the charged Fermi gas. These are well-known within Lindhard
theory \citep{Lindhard_1954} in the self-consistent field (or random-phase) approximation.
By matching the long-wavelength expansion of these functions, extended to take into
account collisions \citep{Kliewer_1969a, Mermin_1970, Conti_1999}, 
we find a kinematic shear viscosity that scales, below the collision rate $1/\tau$,
with $v_{\rm F} \ell = \ell^2 / \tau$. This is formally a diffusion coefficient comparable 
in magnitude to
the magnetic one $D\bulk{m} = \lambdabar_p^2 / \tau$ because of the coincidence
between the (reduced) plasma wavelength $\lambdabar_p = c / \OmegaP = 22\,{\rm nm}$ 
and the mean free path $\ell$.

\begin{figure}[bth]
\centerline{%
\includegraphics*[width=0.7\textwidth]{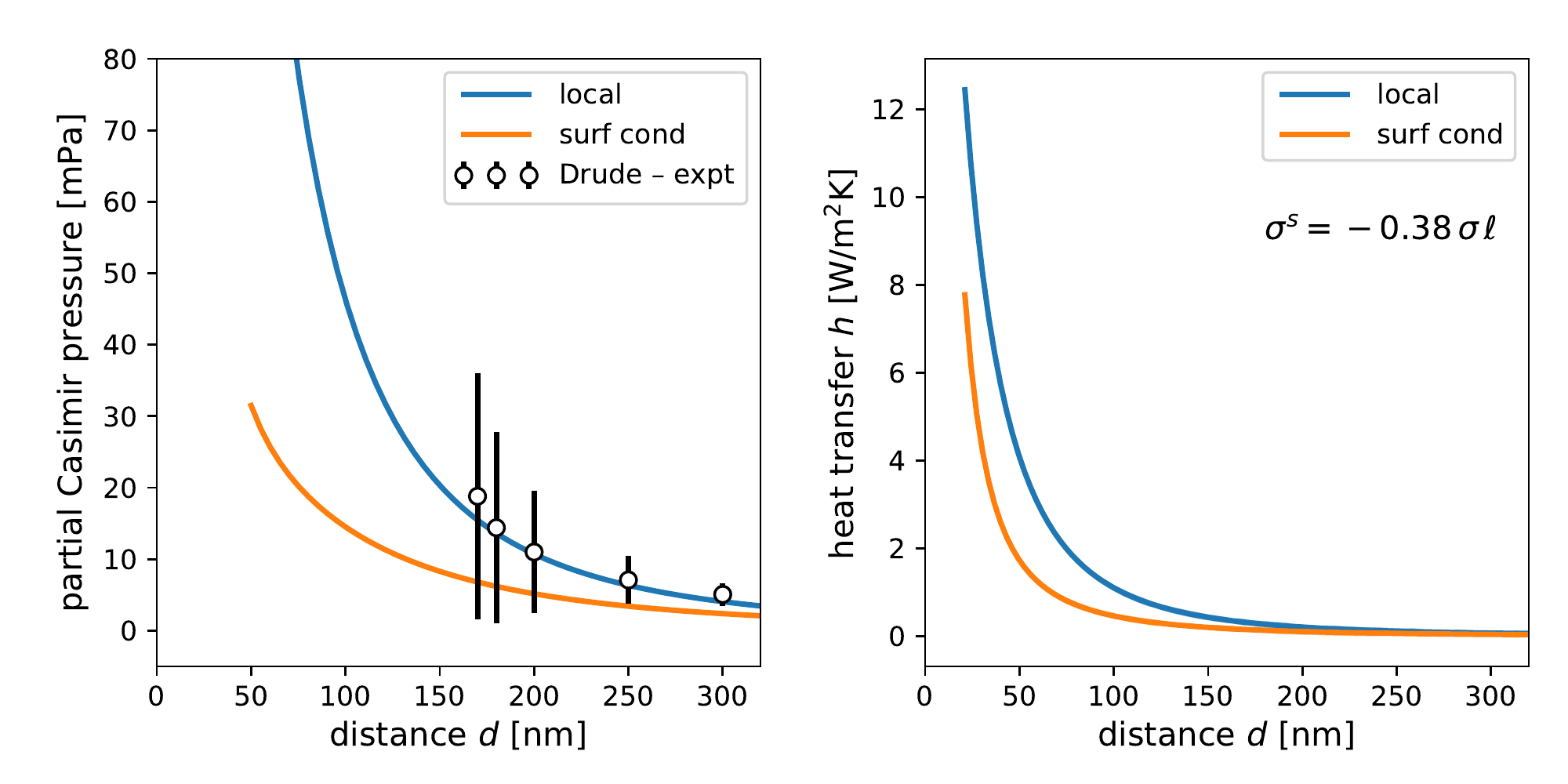}
}
\caption[]{(left) Partial Casimir pressure (only s-polarized waves are included) between
two thick plates as a function of distance $d$. 
{The solid lines give the \emph{repulsive} thermal contribution (the
$T = 0$ limit subtracted) for two response functions of the metal. 
The open black symbols with error bars give the absolute difference between the observed 
pressure and the full theoretical calculation based on the local Drude approximation 
(data from \citet{Decca_2005}).}
(right) Heat transfer coefficient $h(d, T) = S(d, T + \Delta T, T) / \Delta T$
[see Eqs.\,(\ref{eq:heat-transfer-evan}), (\ref{eq:heat-transfer-prop})],
as the temperature difference $\Delta T \to 0$. The results are dominated by 
evanescent waves with $k \ge \omega/c$, propagating waves giving a negligible contribution.
Local: formulas~(\ref{eq:Fresnel_TE}) for reflection coefficient $r_s$;
surf cond: model of Sec.\,\ref{s:Bedeaux-Vlieger} with surface conductivity $\surconduct$.
Material parameters for gold as given in the main text,
temperature $T = 300\,{\rm K}$.
}
\label{fig:map-Casimir}
\end{figure}

An overview of our results is shown in Fig.\,\ref{fig:map-Casimir}
where the Casimir pressure (left) and the heat transfer (right) due to s-polarized modes
is plotted. 
{These modes give a sizeable
thermal correction to the Casimir force \citep{Bostrom_2000b, Torgerson_2004, Intravaia_2009a}
and reduce the attraction prevailing between ideal reflectors.
The p-polarisation does not contribute because of the efficient charge build-up at the metal surfaces
(in the infrared, the dielectric screening is $1/|\varepsilon\bulk{m}| \ll 1$). 
Note in particular that the s-polarized modes alone account for nearly 
the entire difference between the measured Casimir force and the theory based 
on the local Drude approximation (open dots with error bars \citep{Decca_2005}).
In the no-slip model, their
repulsive contribution drops significantly compared to the local Drude model, 
so that the difference to observations gets smaller \citep{Klimchitskaya_2020b}.}
The radiative heat current in Fig.\,\ref{fig:map-Casimir}\,(right) is also reduced, but
the data are well below the levels observed by \citet{Kloppstech_2017, Cui_2017a}.

The outline of the paper is as follows. In Sec.\,\ref{s:Casimir-Fresnel}, we motivate
the formulas for the Casimir pressure between metallic plates at nonzero temperature
and for the radiative heat transfer. 
After an introduction to the hydrodynamic approximation (Navier-Stokes equation) 
in Sec.\,\ref{s:intro-hydrodynamics}, 
we solve the reflection/transmission problem at a conducting surface
using the no-slip boundary condition 
(Sec.\,\ref{s:hydrodynamic-r|t-solution})
and a modified surface current density (Sec.\,\ref{s:Bedeaux-Vlieger}).
A discussion of the reflection coefficients and the impact on the Casimir pressure
is given in Sec.\,\ref{s:discussion}.
Appendix~\ref{a:Lindhard-Conti} presents 
the derivation of the viscosity for conduction electrons based on the wave vector-
and frequency-dependent longitudinal and transverse dielectric functions.

\section{Casimir pressure and boundary conditions}
\label{s:Casimir-Fresnel}

\begin{figure}[bt]
\centerline{\includegraphics[width=0.3\textwidth]{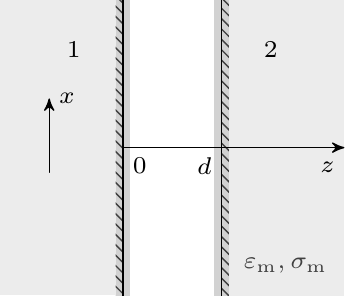}}
\caption[]{Simple system setup: two metallic plates (1 and 2) separated by a vacuum gap 
of distance $d$. In the sub-surface regions (hatched), the hydrodynamic and local models
give different profiles for the current density.}
\label{fig:System_setup}
\end{figure}

The famous formula by Casimir for the force per unit area between two ideally
reflecting plates (Fig.\,\ref{fig:System_setup}) reads
\begin{equation}
\frac{ F_C }{ A } = - \frac{ \pi^2 }{ 240 } \frac{ \hbar c }{ d^4 }
\label{eq:Casimir-1948b}
\end{equation}
where $d$ is the distance and the negative sign denotes an attractive force.
This formula ignores the physical
properties of the plates, although its derivation requires that they become transparent
in the far UV -- to regularise the UV divergent vacuum energy.

A more complete description is provided by the theory of dispersion forces.
It describes electromagnetic fluctuations, 
both in the quantum
and thermal regime, in and between macroscopic bodies and the ensuing interactions.
If we restrict the discussion to distances $d \gg 1\,\text{\AA}$,
it seems appropriate to use a continuum description and to describe the objects with the
help of material equations, using the framework of macroscopic electrodynamics.
This has been developed over the last twenty years into macroscopic quantum electrodynamics
\citep{Rytov_vol3, Buhmann_BookI, Buhmann_BookII, Volokitin_2017book, Sernelius_2018book}. 
Its basic idea is that the macroscopic response functions
also determine the strength of the fluctuations produced by the bodies. 
By the very construction of this approach, there is
no distinction to be made between virtual or real fields: 
the body's material is responding to an external field.

We focus on the simple geometry of two thick parallel plates a distance $d$ apart, 
with the $z$-axis normal to the surfaces. 
They are kept at temperature $T$ so that 
for a given (angular) frequency $\omega$, the mean energy per photon mode is
given by $\frac12 \hbar \omega \coth( \hbar \omega / 2 k_B T) = 
\hbar \omega [ \frac12 + \bar n( \omega, T ) ]$ defining the Bose-Einstein distribution
$\bar n$. Due to translational and rotational symmetry parallel 
to the plates, the
electromagnetic modes may be labelled by a two-dimensional $k$-vector 
${\bf k} = (k_x, k_y)$. Between the plates (vacuum), these modes vary with the wave vector
\begin{equation}
k_z = \left\{
\begin{array}{ll}
\sqrt{ (\omega/c)^2 - k^2 } & \mbox{for } k \le \omega/c \quad \mbox{(propagating mode)}
\\[1ex]
{\rm i} \sqrt{ k^2 - (\omega/c)^2 } = {\rm i} \kappa & \mbox{for } k \ge \omega/c
\quad \mbox{(evanescent mode)}
\end{array}
\right.
\label{eq:normal_wavevector}
\end{equation}
In the second case, the modes are called evanescent, and they 
are localised to the vicinity of their sources. 
(\citet{Klimchitskaya_2020b} use the words ``on-shell'' (``off-shell'') for 
propagating (evanescent) modes, respectively.)
There are two
transverse polarisations, usually called p (or TM) and s (TE). When a wave with
polarisation $\pol$ is incident on the metal plate, it is reflected with amplitude
$r_\pol = r_\pol(k, \omega)$. Multiple reflections between plate~1 and~2 can be represented 
by a geometric series
\begin{equation}
1 + r_{1\pol} r_{2\pol} \,{\rm e}^{ 2 {\rm i} k_z d } + \ldots =
\frac{ 1 }{ 
1 - r_{1\pol} r_{2\pol} \,{\rm e}^{ 2 {\rm i} k_z d } 
}
\label{eq:cavity-loop-factor}
\end{equation}
At each reflection, a propagating photon imparts a recoil momentum of order $\hbar k_z$ 
onto the plate. Taking into account the reflection amplitudes and summing over all photon modes 
and their thermal occupation numbers, the electromagnetic stress normal to the surfaces
yields the Lifshitz formula for the Casimir force per unit area \citep{Lifshitz_1956}
\begin{equation}
P(d,T) = 
\mathop{\rm Re} 
\int\limits_0^\infty\!\frac{d\omega}{2\pi} \coth\frac{\hbar\omega}{2k_BT}
\int\limits_{\sf L}\!\frac{k_z \, dk_z }{2\pi} 
2 \hbar k_z \sum_{\pol \,=\, {\rm s,p}} 
\frac{ r_{1\pol}r_{2\pol} \,{\rm e}^{2{\rm i} k_z d} 
    }{ 
       1- r_{1\pol}r_{2\pol} \,{\rm e}^{2{\rm i} k_z d} 
     }
\label{eq:Lifshitz56}
\end{equation}
This contains also the contribution of evanescent waves via the ${\sf L}$-shaped
path of the $k_z$ integral: it runs along the imaginary axis from ${\rm i}\infty$ to 
$0$ and then to $\omega/c$ (recall the convention: negative $P$ gives an
attractive force).

Formula~(\ref{eq:Lifshitz56}) is not suitable for calculating the pressure
because its zero-temperature limit is plagued by the rapid oscillating factor
${\rm e}^{2{\rm i} k_z d}$ at high frequencies (along the real ``leg'' of the $k_z$-integral).
(The integral is physically cut off around the the plasma frequency where 
tabulated optical data rather than the Drude permittivity must be used.)
Lifshitz shifted the $\omega$-integration in the complex frequency plane
to the imaginary axis $\omega = {\rm i} \xi$ which is possible because the integrand
is built from response functions that are analytic in the upper half plane.
The $k_z$-integral is then
taken from ${\rm i}\infty$ to ${\rm i}\xi/c$ so that all exponentials become real
and decay at large $\xi$. 
For finite temperatures, the integration is replaced by a summation 
over the Matsubara frequencies $\omega = {\rm i}\xi_n = 2\pi {\rm i} n k_BT / \hbar$, 
the poles of $\coth( \hbar \omega / 2k_BT)$, with the term $n = 0$ counting only one half:
\begin{equation}
\mathop{\rm Re}
\int\limits_0^\infty\!\frac{d\omega}{2\pi} \coth\frac{\hbar\omega}{2k_BT} f( \omega )
\mapsto
\frac{ k_BT }{ \hbar } \sideset{}{'}\sum_n f( {\rm i} \xi_n )
\label{eq:map-to-Matsubara-sum}
\end{equation}
We have used here that the integrand $f( \omega )$ becomes a real function along the
imaginary axis. 

Similar considerations have led \citet{Polder_1971} and \citet{Loomis_1994} 
to a formula for the heat current between two planar bodies of temperatures $T_1 > T_2$ 
separated by a vacuum gap of width $d$ \citep{Volokitin_2017book}. 
For small gaps, the important contribution 
comes from evanescent waves, i.e. imaginary $k_z = {\rm i} \kappa$:
\begin{equation}
    S_{\rm evan} (d, T_i) = \int\limits_0^\infty \! \frac{ d\omega }{ 2\pi }\,
    \hbar \omega \Big[ \bar n( \omega, T ) \Big]^{T_1}_{T_2}
     \int\limits_0^\infty \frac{\kappa\, d\kappa }{2\pi}  
    \sum_{\pol \,=\, {\rm s,p}} 
    \frac{ 4 \mathop{\rm Im}(r_{1\pol}) \mathop{\rm Im}(r_{2\pol}) \,{\rm e}^{-2\kappa d} 
    }{ 
       \left| 1 - r_{1\pol}r_{2\pol} \,{\rm e}^{-2\kappa d} \right|^2
    }
\label{eq:heat-transfer-evan}
\end{equation}
where $\big[ \bar n( \omega, T ) \big]^{T_1}_{T_2} 
= \frac12 \coth(\hbar \omega / 2k_B T_1) - \frac12 \coth(\hbar \omega / 2k_B T_2)
= \bar n( \omega, T_1 ) - \bar n( \omega, T_2 )$
is the difference of Bose distributions. 
The contribution from real $k_z$ looks a bit different,
\begin{equation}
    S_{\rm prop} (d,T_i) = \int\limits_0^\infty \! \frac{d\omega}{2\pi}
\hbar \omega \Big[ \bar n( \omega, T ) \Big]^{T_1}_{T_2}
\int\limits_0^{\frac{\omega}{c}} \frac{k_z\, dk_z }{2\pi}  
\sum_{\pol \,=\, {\rm s,p}} 
\frac{ \big( 1 - \left|r_{1\pol} \right|^2 \big) \big( 1 - \left|r_{2\pol} \right|^2 \big)
}{ 
    \left| 1 - r_{1\pol}r_{2\pol} \,{\rm e}^{2{\rm i} k_z d}   \right|^2 
} \, .
\label{eq:heat-transfer-prop}
\end{equation}
Physical properties of the reflection coefficients, namely energy conservation for real $k_z$
($|r_{i\pol}|^2 \le 1$) 
and passivity for imaginary $k_z$ ($\mathop{\rm Im} r_{i\pol} \ge 0$), ensure that the heat
current is always oriented from hot to cold, consistent with the Second Law of thermodynamics.
Note that in this approach, the concept of temperature has shifted from the field itself
to its sources, namely currents and charges in the two bodies. A detailed discussion when
the simple picture of two uniform temperatures $T_1 \ne T_2$ is applicable, has been given
by \citet{Eckhardt_1982}. The basic idea is that the material's heat capacity and thermal
conductivity are sufficiently large so that the absorption of electromagnetic energy does not
change its temperature. Similar arguments have been used to model transport in 
semiconductors at high fields \citep{Ancona_1995}.
Additional bodies in thermal contact are obviously also instrumental 
in maintaining the non-equilibrium setting.

In the following, our focus will be on the temperature-dependent part of the
Casimir pressure and the radiative heat transfer. This is why
we do not use the Matsubara sum: the thermal
correction would be hidden in the difference between sum and integral 
[see Eq.\,(\ref{eq:map-to-Matsubara-sum})]. Using the argument
principle rather than the Euler-MacLaurin formula to evaluate that difference, brings
us back to the real-frequency integral~(\ref{eq:Lifshitz56}). 
This makes one essential difference with respect to \citet{Klimchitskaya_2020b} 
where a modified surface response was also proposed, but the focus was
on the behaviour of the zero'th Matsubara frequency $\xi_0 = 0$. Further comparison
to that paper will be drawn in the Conclusion.
Since we also focus
on distances $d$ much smaller than the thermal wavelength $\lambdabar_T$, the
$k_z$-integral is dominated by its imaginary leg (evanescent modes), thus providing
numerically tractable expressions. A typical scale for the imaginary wavenumber
is set by the inverse distance $\kappa \sim 1/d$.

It remains in the following to analyse the reflection coefficients. 
If the plates are characterized by a local dielectric function $\varepsilon\bulk{m}(\omega)$ 
(a conductivity $\sigma\bulk{m}(\omega)$),
the Fresnel formulas can be used
\begin{align}
r_s &= \frac{k_z - {\rm i} \kappa\bulk{m}}{k_z + {\rm i} \kappa\bulk{m}}
\qquad
& r_p = \frac{\varepsilon\bulk{m} k_z - {\rm i} \kappa\bulk{m}}{\varepsilon\bulk{m} k_z + {\rm i} \kappa\bulk{m}} 
\label{eq:Fresnel_TE}
\end{align}
where 
\begin{equation}
\kappa\bulk{m} = \sqrt{ k^2 - (\omega/c)^2 \varepsilon\bulk{m} }
\label{eq:def-kmz}
\end{equation}
is the decay constant inside the metal.
(We take $\mathop{\rm Re} \kappa\bulk{m} \ge 0$.)
These expressions arise from the dispersion relation
$k^2 - \kappa\bulk{m}^2 = (\omega/c)^2 \varepsilon\bulk{m}$
in the metal and by matching the tangential components of the electric and magnetic
fields at the vacuum-metal interface. In the following section, we derive a 
generalisation of these expressions using a hydrodynamic picture where the conduction
electrons are modelled as a charged, visco-elastic medium.

\section{Visco-elastic electron dynamics}

\subsection{Bulk}
\label{s:intro-hydrodynamics}

The key concepts for a hydrodynamic description are the density $n$ and the 
velocity field ${\bf v}$ for
carriers with mass $m$ and charge $e$. The dynamics of the latter is given by 
$n m ( \partial_t + {\bf v} \cdot \nabla) {\bf v} = {\bf f}$ with the Navier-Stokes force density
\citep{landau1987fluid}
\begin{eqnarray}
	{\bf f} & = & n e ({\bf E} + {\bf v} \times {\bf B})
	- \frac{ n m }{ \tau } {\bf v}
	- m \beta^2 \nabla n
	+ n m \eta \nabla^2 {\bf v}
	+ n m \zeta' \nabla (\nabla \cdot {\bf v})
	\label{eq:Navier-Stokes}
\end{eqnarray}
Here, $1/\tau$ is the Drude scattering rate (see discussion in Appendix~\ref{a:Lindhard-Conti}), 
the compressibility is expressed via the velocity
$\beta$, and the kinematic shear and bulk viscosities are $\eta$, $\zeta$
with $\zeta' = \zeta + \frac13\eta$.
(This force density can be interpreted as a gradient expansion, 
assuming that $n$ and ${\bf v}$ vary on large scales only. The hydrodynamic description
does not resolve a microscopic scale like the Fermi wavelength.)

We are interested in the linear response of conduction electrons to the electric field
that splits naturally into a longitudinal ($L$) and transverse ($T$) part
${\bf E} = - \nabla \phi - \partial_t {\bf A}$ (Coulomb gauge: $\nabla \cdot {\bf A} = 0$).
Using the equation of continuity $\partial_t n + \nabla \cdot( n {\bf v} ) = 0$,
assuming all fields to evolve at a given frequency with $\exp( - {\rm i}\omega t)$, 
and dropping second-order terms from the Navier-Stokes equation~(\ref{eq:Navier-Stokes}),
we obtain
\begin{eqnarray}
- {\rm i} \omega {\bf v}_{L} &=& - \frac{e}{m} \nabla \phi - \frac{ {\bf v}_{L} }{ \tau } 
+ \frac{ {\rm i} }{ \omega } \left( \beta^2 - {\rm i}\omega \zeta' \right) \nabla (\nabla \cdot {\bf v}_{L}) + \eta \nabla^2 {\bf v}_{L}
\label{eq:longitudinal-linear-response}
\\
- {\rm i} \omega {\bf v}_T &=& {\rm i} \omega \frac{e}{m} {\bf A} - \frac{ {\bf v}_T }{ \tau } 
+ \eta \nabla^2 {\bf v}_T
\label{eq:transverse-linear-response}
\end{eqnarray}
The complex combination $\beta^2 - {\rm i}\omega \zeta'$ can be interpreted as a dynamic
modulus, using the language of visco-elastic media \citep{Conti_1999}.
In a homogeneous medium where the fields vary with the wave vector ${\bf q}$, these equations 
produce the longitudinal and transverse conductivities in the hydrodynamic approximation
according to 
${\bf j}_{L,T} = n_e e \, {\bf v}_{L,T} = \sigma_{L,T}( {\bf q}, \omega ) {\bf E}_{L,T}$:
\begin{eqnarray}
\sigma_{L}({\bf q}, \omega) &=& 
\frac{ \sigma_0 }{
1 - {\rm i} \omega \tau + ({\rm i} \beta^2 / \omega + \zeta' + \eta) \tau q^2 }
\label{eq:hdyn-longitudinal-sigma}
\\
\sigma_T({\bf q}, \omega) &=& 
\frac{ \sigma_0 }{
1 - {\rm i} \omega \tau + \eta \tau q^2 }
\label{eq:hdyn-transverse-sigma}
\end{eqnarray}
where $\sigma_0 = n_e e^2 \tau / m$ is the DC conductivity 
and $n_e e$ the equilibrium charge density. Note from the poles of these expressions 
for $\tau \to \infty$ how $\beta$ determines the speed of longitudinal sound waves, 
while the transverse current behaves in a diffusive way with diffusion constant $\eta$.

\begin{figure}[tbh]
\centerline{\includegraphics*[width=0.55\textwidth]{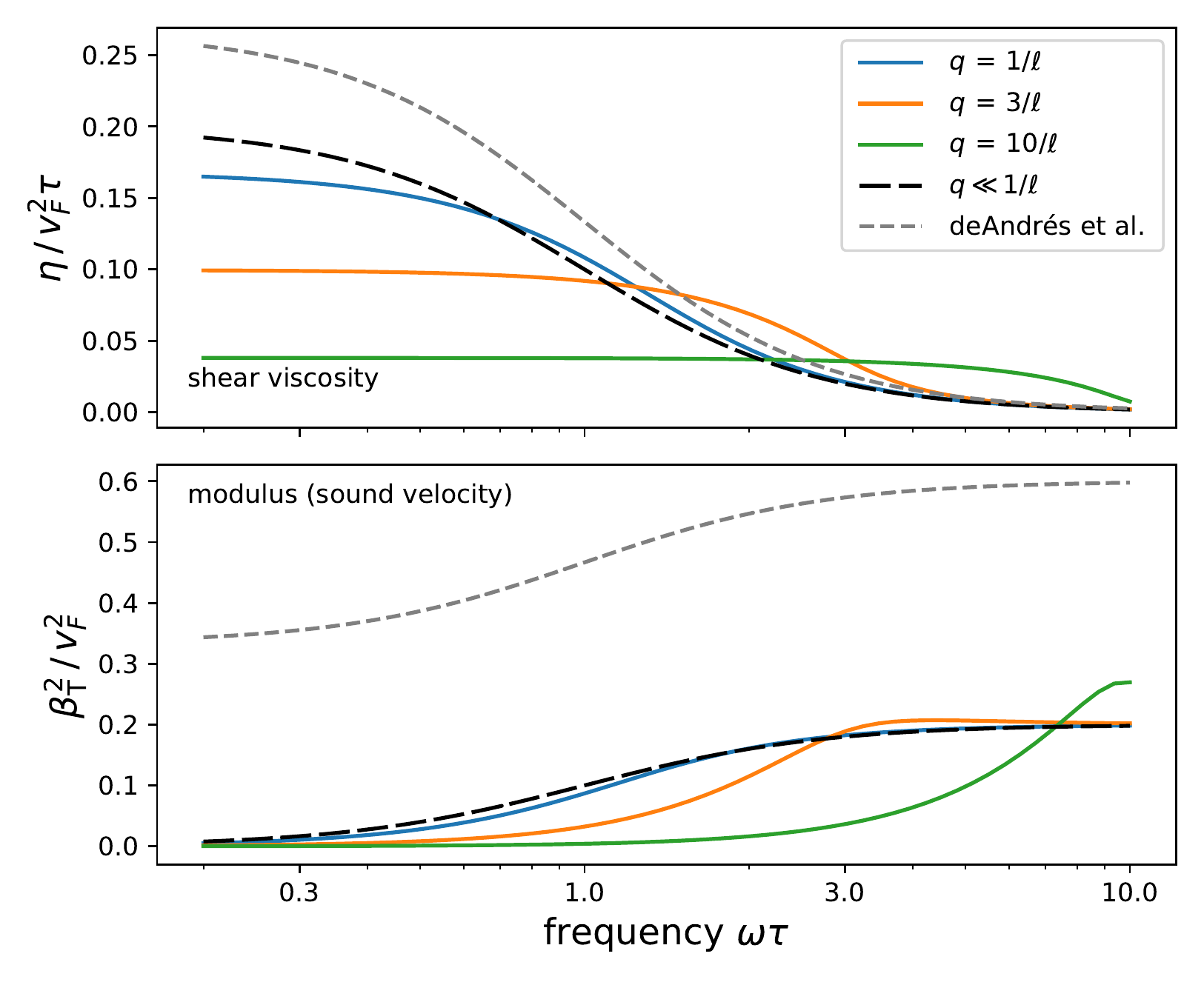}}
\caption[]{Shear viscosity (top) and modulus (bottom) for the Fermi gas with collisions,
based on the transverse Lindhard dielectric function modified according to \citet{Conti_1999}.
We plot the kinematic viscosity $\mathop{\rm Re}\eta(\omega)$ and the square of
the shear wave sound velocity $\beta_{T}^2 = \omega\mathop{\rm Im}\eta(\omega)$.
The thick black dashed line corresponds to the hydrodynamic
limit (small $q$). The gray dashed line is the result of \citet{deAndres_1986}
(see Appendix~\ref{a:deAndres-et-al}). 
The solid colored lines are obtained 
from the inverse conductivity $\sigma_0/\sigma_{T}({\bf q}, \omega)$ computed 
according to \citet{Kliewer_1969a} and \citet{Conti_1999}, by subtracting the local limit 
$1 - {\rm i}\omega \tau$ and dividing by $q^2\tau$
[see Eq.\,(\ref{eq:hdyn-transverse-sigma})
and Appendix~\ref{a:transverse-epsilon-from-Lindhard}].
The kinks appearing at $\omega \sim v_{\rm F} q$ signal the onset of Landau damping
(creation of electron-hole pairs).}
\label{fig:shear-viscosity}
\end{figure}

The expressions~(\ref{eq:hdyn-longitudinal-sigma}, \ref{eq:hdyn-transverse-sigma})
provide a framework to actually find the hydrodynamic parameters 
for the electron fluid. In this paper, we focus on the seminal results of
\citet{Lindhard_1954} for the conductivities in the self-consistent field (or
random-phase) approximation, leaving a detailed study of exchange-correlation
effects for later work (see \citet{Conti_1999} for this more general approach).
We incorporate collisions into the Lindhard functions in such a way that
charge excitations relax to local equilibrium set by the electrochemical potential
$E_{\rm F} + e \phi$ and their static limit is correctly reproduced
\citep{Kliewer_1969a, Mermin_1970, Conti_1999}. 
The Lindhard conductivities
are not restricted by the hydrodynamic approximation and can be expanded for 
small $q$-vectors. As outlined in Appendix~\ref{a:Lindhard-Conti}, this procedure gives
\begin{eqnarray}
\beta^2 - {\rm i} \omega \left[\zeta'(\omega) + \eta(\omega)\right]
&=& v_{\rm F}^2 
\frac{1/3 - 3i\omega \tau/5}{1 - i\omega \tau} 
\label{eq:Halevi-beta}\\
\eta(\omega) &=& 
\frac{v_{\rm F}^2 \tau}{5 \left( 1 - i\omega \tau \right)} 
\label{eq:Lindhard-Conti-eta}
\end{eqnarray}
The expression~(\ref{eq:Halevi-beta}) recovers the longitudinal speed of sound 
derived by \citet{Halevi_1995}. 
Its real part crosses over from $\beta = v_{\rm F}/\sqrt{3}$ at low
frequencies to $\sqrt{\frac35} v_{\rm F}$ at high frequencies (isentropic limit).
Its imaginary part is attributed here to the viscosities $\zeta, \eta$ of the electron fluid.
The shear viscosity~(\ref{eq:Lindhard-Conti-eta}) 
turns out to be larger than the quantum scale $\hbar/m$
and is plotted in Fig.\,\ref{fig:shear-viscosity} (limit $q \ell \ll 1$, black dashed lines).
Its low-frequency limit is set by the diffusion constant $\tfrac15 v_{\rm F}^2 \tau 
= \tfrac15 v_{\rm F} \ell$.
The lower part of the Figure gives the corresponding shear velocity $\beta_T$, defined from
$\beta_T^2 = \mathop{\rm Re}[-{\rm i}\omega \eta(\omega)]$.
It scales $\beta_{T}^2 \sim \omega^2$ at low frequencies, 
consistent with the picture that a liquid does not support low-frequency shear waves.
Around the collision frequency, the response to shear changes from viscous to elastic
and $\beta_{T} \approx v_{\rm F}/\sqrt{5}$ for $\omega \gg 1/\tau$.
The gray dashed lines correspond approximately to the longitudinal elastic parameters 
(see discussion
of \citet{deAndres_1986} in Appendix~\ref{a:deAndres-et-al}).
The other curves in Fig.\,\ref{fig:shear-viscosity} illustrate corrections beyond 
hydrodynamics that appear when the dimensionless 
parameter $q v_{\rm F} / \left( \omega + {\rm i}/\tau \right) \sim 1$. They become relevant
on scales shorter than the mean free path, i.e., $q \ell > 1$.

We make the following curious observation from Eqs.\,(\ref{eq:Halevi-beta},
\ref{eq:Lindhard-Conti-eta}): when the longitudinal speed of sound $\beta$ and the bulk
viscosity $\zeta$ are computed by subtracting the part involving the shear viscosity $\eta$,
one obtains $\beta^2 - {\rm i} \omega \zeta = \frac13 v_{\rm F}^2$, a real constant.
The bulk viscosity $\zeta$ thus vanishes (as also mentioned by \citet{Conti_1999}).
The dispersion of longitudinal sound waves found by \citet{Halevi_1995} 
originates entirely from the nonzero shear modulus $\tfrac43 \mathop{\rm Re}[-{\rm i}\omega \eta(\omega)]$.

We finally note a close coincidence of parameters. The penetration of s-polarised evanescent 
fields into a metal follows the decay constant
$\kappa\bulk{m} \approx [k^2 - {\rm i} \omega \mu_0 \sigma_0/(1-{\rm i}\omega\tau)]^{1/2}$ 
where the DC conductivity $\sigma_0$ can be expressed 
as a diffusion constant $1/\mu_0\sigma_0 = \lambdabar_p^2 / \tau$ that governs the
spatio-temporal behaviour of low-frequency magnetic fields \citep{jackson2014klassische}.
The kinematic shear viscosity $\eta$ has also the dimension of a diffusion coefficient
[see Eq.(\ref{eq:hdyn-transverse-sigma})] $\eta \sim v_{\rm F}^2 \tau = \ell^2 / \tau$.
For the noble metal parameters, we have $\ell \sim \lambdabar_p$, and the diffusive behaviour
of both types overlaps in space.
The only way to formally isolate the local Drude model is the ``dirty limit'' 
where $\tau \to 0$ at fixed $\lambdabar_p$, $v_{\rm F}$. A discussion of the opposite
case, that is typical for low temperatures, is provided by 
\citet{Intravaia_2010c, Reiche_2020a}.

\subsection{Sub-surface region}
\label{s:hydrodynamic-r|t-solution}

We now apply the Navier-Stokes equations to the response of a metallic half-space
to an s-polarized field and compute the reflection amplitude $r_s$. We assume
that all fields vary $\sim \exp{\rm i}( k x - \omega t )$ with the $x$-axis parallel
to the surface and the metal occupying the region $z \ge 0$. It is easy to see that
the s-polarisation gives transverse fields \emph{senkrecht} (orthogonal) to the $xz$-plane, 
we denote by $v = v(z)$ and $A = A(z)$ the corresponding components of ${\bf v}_{T}$ 
and ${\bf A}$. The no-slip boundary condition $v(0) = 0$ thus pertains to the
tangential velocity of the electron fluid, while $A$, proportional to the tangential
electric field, is actually continuous across the surface. 
This rule has the advantage of not
needing a dimensional parameter (apart from the bulk viscosity fixed from the bulk
behaviour). It nevertheless provides an ``additional boundary condition'' in the language
of optics in spatially dispersive media \citep{DresselGruenerBook}. 
The approach of \citet{Klimchitskaya_2020b} is very different since they modify the
${\bf q}$-dependence of the bulk dielectric function.
Compared to Eqs.\,(\ref{eq:hdyn-longitudinal-sigma}, \ref{eq:hdyn-transverse-sigma}), 
their corrections are linear in $k$ rather than quadratic, the anisotropy being justified
by the presence of the surface. Our approach is quite the opposite, 
since an explicit boundary condition enters into the description of the surface, 
while (deep) inside the metal, the bulk dielectric functions are applied,
keeping their spatial dispersion within the scope of hydrodynamics.

Eq.\,(\ref{eq:transverse-linear-response}) and the Maxwell equations yield 
the following equations of motion
\begin{eqnarray}
\frac{ {\rm d}^2 v }{ {\rm d}z^2 } &=& 
\left( 
\frac{ 1 - {\rm i}\omega \tau }{ \eta \tau } + k^2 \right) v 
- \frac{ {\rm i}\omega }{ \eta }\frac{e A}{m}
\label{eq:transverse-v-half-space}
\\
\frac{ {\rm d}^2 A }{ {\rm d}z^2 } &=&
- \frac{ m }{ e \, \lambdabar_p^2} v + \kappa\bulk{b}^2 A 
\label{eq:A-half-space}
\end{eqnarray}
where we used the link between electron
density and plasma wavelength $\lambdabar_p = c/\Omega_p$ to re-write the 
current density $\mu_0 n_e e v$ [first term of Eq.\,(\ref{eq:A-half-space})].
In $\kappa\bulk{b}^2 = k^2 - \varepsilon\bulk{b}(\omega/c)^2$, the displacement current will give
a negligibly small contribution.
Since the fields must decay deep into the bulk metal, 
this system can be solved with the \emph{Ansatz} 
\begin{equation}
v(z) = v_1 \, {\rm e}^{ - \kappa_1 z } + v_2 \, {\rm e}^{ - \kappa_2 z }
\label{eq:Ansatz-velocity-profile}
\end{equation}
and similar for $A$. The decay constants are given by
\begin{eqnarray}
	\kappa_{1,2}^2 &=& 
	\frac{ 1 }{ 2 } 
	\left( \frac{ 1 - {\rm i} \omega \tau }{ \eta \tau } + k^2 + \kappa\bulk{b}^2
	\right)
	\pm
	\frac{ 1 }{ 2 } 
	\left[
	\Big( \frac{ 1 - {\rm i} \omega \tau }{ \eta \tau } + k^2 - \kappa\bulk{b}^2 \Big)^2 
	+ \frac{ 4 {\rm i} \omega }{ \eta \lambdabar_p^2 }
	\right]^{1/2}
	\label{eq:result-kappa-12}
\end{eqnarray}
The ratio between the eigenmode amplitudes $v_{l}$,
$A_{l}$ ($l = 1,2$) is $m v_{l} / e A_{l} = (\kappa\bulk{b}^2 - \kappa_{l}^2) \lambdabar_p^2$
[Eq.\,(\ref{eq:A-half-space})].
It is essential to have two decay modes here, otherwise the boundary
conditions $v( 0 ) = 0 = v( \infty )$ would make the velocity vanish everywhere. For typical
good conductors and $\omega \tau \sim 1$, the two terms
under the root are comparable. If the mean free path is not resolved, $\kappa_1 \approx
(1 - {\rm i}\omega \tau)^{1/2}/\ell$ diverges (thin boundary layer) and
$\kappa_2 \approx \kappa\bulk{m} 
= (\kappa\bulk{b}^2 - {\rm i} \omega \mu_0 \sigma_0/(1 - {\rm i}\omega\tau))^{1/2}$ 
is the decay constant in the local Drude model and the Fresnel equation~(\ref{eq:Fresnel_TE}).

The no-slip boundary condition fixes from Eq.\,(\ref{eq:Ansatz-velocity-profile})
the ratio $v_1 = - v_2$,
so that only one free parameter remains. It is fixed by
the amplitude of the field incident from the vacuum side. A convenient quantity
is the ratio $Z = -A / (dA / dz)$ (a length) evaluated at the surface. 
Since $E_y = {\rm i}\omega A$ is 
a tangential electric field and $B_x = - d A / dz$ a tangential magnetic field,
the ratio $Z$ is actually proportional to the surface impedance of the metallic half-space. 
A quick calculation gives
\begin{equation}
Z 
 = \frac{ \kappa_2 + \kappa_1
   }{ \kappa\bulk{b}^2 + \kappa_1 \kappa_2 }
\label{eq:}
\end{equation}
The ``impedance'' $Z$ matches with the incident and reflected fields on the vacuum side, 
$A(z) = A_{0} \left( {\rm e}^{ {\rm i} k_z z } + r_s \, {\rm e}^{ - {\rm i} k_z z } \right)$. 
This yields the beautiful
formula for the s-polarised reflection amplitude
\begin{equation}
r_s = \frac{ k_z - {\rm i}/Z }{ k_z + {\rm i}/Z } \approx 
\frac{ ( k_z - {\rm i}\kappa_2)({\rm i}\kappa_1 - k_z) }{ ( k_z + {\rm i}\kappa_2)({\rm i}\kappa_1 + k_z) }
\,,
\label{eq:}
\end{equation}
the main result of this section. (The second form becomes exact for $\varepsilon\bulk{b} = 1$.)
In the local limit (vanishing viscosity), $\kappa_1$
diverges [Eq.\,(\ref{eq:result-kappa-12})], 
and $r_s$ goes into the Fresnel formula~(\ref{eq:Fresnel_TE}). 
Corrections to this thus depend on the ratio $\kappa_1 / k_z$.

\begin{figure}[bth]
\centerline{%
\includegraphics*[width=0.65\textwidth]{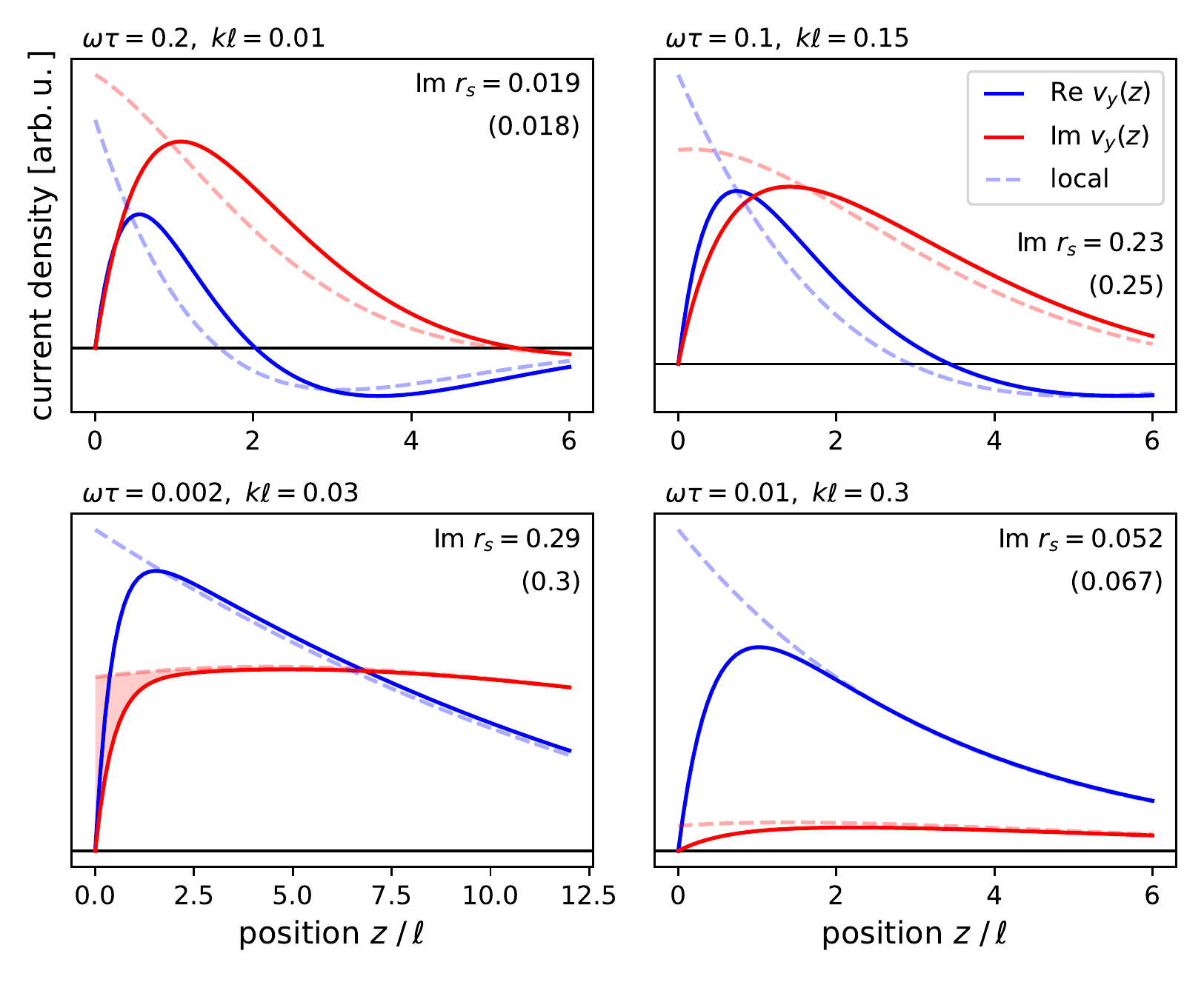}%
}
\caption[]{Sub-surface distribution of current density
for different choices
in the $k,\omega$-plane (marked by white dots in Fig.\,\ref{fig:map_both}).
The shaded area in the bottom left illustrates the concept of an excess current
(Sec.\,\ref{s:Bedeaux-Vlieger}).
The numbers for $r_s$ correspond to the hydrodynamic no-slip model and the local
Drude model (in parentheses). The data are based on a unit amplitude (real-valued)
electric field incident from the vacuum side (left).
}
\label{fig:current-profile}
\end{figure}

The hydrodynamic description is illustrated by the results in Fig.\,\ref{fig:current-profile}
where the current profile $v(z)$ is shown for different choices of parameters. 
The values of the reflection coefficients are also given and compared to the local
(Drude-Fresnel) result.

\subsection{Reduced boundary layer conductivity}
\label{s:Bedeaux-Vlieger}

It is remarkable in Fig.\,\ref{fig:current-profile}\,(bottom left) how the current distribution 
is ``missing'' a sub-surface sheet a few $\ell$ thick,
when comparing to the local model that does not
apply the no-slip boundary condition (light dashed lines). We outline in this section
how this can be included into a modified boundary condition for the electromagnetic fields,
using the excess field technique developed by \citet{Bedeaux_2002}. This is actually a
paradigmatic example of boundary layer approximations or multiple-scale expansions
\citep{NayfehBook,BenderOrszagBook}. The response of the charge density at a conducting
surface has been recently analyzed in the same spirit by \citet{AsgerMortensen_2021a}.

The excess field approach lumps the details about the behaviour of fields and currents
in the surface region into a small number of response functions. The idea is based 
on a separation
of scales where deviations from a homogeneous bulk material only occur in a thin region
near the surface (sometimes called the selvedge \citep{Sipe_1980b}). For simplicity, we
focus on non-magnetic materials and neglect spatial dispersion in the bulk (far away from
the interface). In the following, we provide a closer look at p-polarized waves
because the calculations are more involved. 

The central concept of an excess field is based on taking the difference between a smooth,
microscopic
field $F(z)$, say, and its approximation $F\bulk{loc}( z )$ that 
extrapolates the local-medium values down to the surface,
\begin{equation}
F(z) - F\bulk{loc}(z) = \begin{cases}
F(z) - F\bulk{v}(z) & \text{for } z < 0 \text{ (vacuum)}
\\
F(z) - F\bulk{m}(z) & \text{for } z > 0 \text{ (metal)}
\end{cases}
\label{eq:def-excess-field}
\end{equation}
\citet{Bedeaux_2002} define
the ``total excess'' as the integral of this difference,
$\excess{F} = \int\!{\rm d}z\, [ F(z) - F\bulk{loc}(z) ]$.
The integral typically converges even
before the variation with depth of $F\bulk{loc}(z)$ sets in, a manifestation of a
separation of length scales.
The excess may still depend on the coordinates $x, y$ in the surface. 
In Fig.\,\ref{fig:current-profile}\,(bottom left), the excess current $j^s = n_e e v^s$ 
would be the shaded area between the no-slip hydrodynamic and the local current profiles.

It now remains to connect the excesses to the fields outside the
surface layer. 
Excess quantities play similar roles as surface charges and currents in macroscopic
electrodynamics and determine
jumps $\left. F \right| := F\bulk{loc}(z \to 0_+) - F\bulk{loc}(z \to 0_-)$
of the coarse-grained electromagnetic fields. For a non-magnetic system and a fixed
frequency, an integration
of the macroscopic Maxwell equations across the interface yields
the boundary conditions \citep{Bedeaux_2002}:
\begin{eqnarray}
\left. {\bf E}_\parallel \right|  & = & \nabla_\parallel \excess[z]{E} \, , \\
\left. D_z \right|  & = & -  \nabla_\parallel \cdot \excess[\parallel]{\bf D} \, ,\\
\left. {\bf B}_\parallel \right|  & = & {\rm i} \mu_0 \omega \, \hat{\bf n} \times \excess[\parallel]{\bf D} \, ,\\
\left. B_z \right|  & = & 0 \, .
\end{eqnarray}
where ${\bf D}$ is the displacement field, 
tangential components carry the index $\Vert$, and $\hat{\bf n}$ is the unit 
normal pointing into the metal.
For simplicity, the surface coordinates $x$, $y$ have been suppressed everywhere.
Note certain jumps that are absent from the ordinary Maxwell boundary conditions.

Finally, material relations specific to the surface are needed to express the surface excesses
by the bulk fields \citep{Bedeaux_2002}.
We focus here on the surface conductivity $\surconduct{}$ and the surface resistivity 
$\surresist{}$, in order to capture the conductive properties of the selvedge.
Introducing $\overline{F} 
= \frac{1}{2} \left( F\bulk{loc}(z \to 0_+) + F\bulk{loc}(z \to 0_-) \right)$
as the average on both sides, one obtains the relations
\begin{eqnarray}
- {\rm i} \omega {\bf D}_\parallel^s  & = & \surconduct{}(\omega)\, \overline{\bf E}_\parallel  \, , \label{eq:excess-boundary-condition-0}
\\
E^s_z   & = & - {\rm i} \omega \surresist(\omega)\, \overline{D}_z  \, ,
\label{eq:excess-boundary-condition}
\end{eqnarray}
The time derivative of ${\bf D}_\Vert^s$ gives, of course, the excess
current tangential to the surface, while $E^s_z$ expresses a potential drop
due to the normal displacement current. 

We proceed to solving the reflection and transmission problem for a p-polarized
wave incident from the vacuum side. A plane-wave \emph{Ansatz} $\sim \exp{\rm i}(k x - \omega t)$
as in Sec.\,\ref{s:hydrodynamic-r|t-solution} leads to complex amplitudes $E_x(z)$
and $E_z(z)$ that away from the surface vary according to
\begin{align}
z < 0: \qquad {\bf E}(z) & = \frac{E_0 c}{\omega} \left[       
\begin{pmatrix}
k_z \\ 0 \\ -k
\end{pmatrix} {\rm e}^{ {\rm i} k_z z} 
+ \begin{pmatrix}
- k_z \\ 0 \\ -k
\end{pmatrix} r_p \, {\rm e}^{-{\rm i} k_z z} \right] \, ,
\\
z > 0: \qquad 
{\bf E}(z) & = \frac{E_0 c}{\omega \sqrt{ \varepsilon\bulk{m} }}      
\begin{pmatrix}
{\rm i}  \kappa\bulk{m} \\ 0 \\ -k
\end{pmatrix} 
t_p\, {\rm e}^{ - \kappa\bulk{m} z} \, , 
\end{align}
where $E_0$ is the amplitude of the incident field,
$n\bulk{m} = \sqrt{\varepsilon\bulk{m}}$ the complex refractive index inside the metal,
and $\kappa\bulk{m}$ given in Eq.\,(\ref{eq:def-kmz}).
The coefficients $r_p$ and $t_p$ are the reflection and transmission amplitudes. 
The corresponding magnetic fields can be obtained by using the relation 
$\omega {\bf B} = {\bf q} \times {\bf E}$ with the Snell-Descartes law giving the wave vectors 
${\bf q}$ on both sides of the surface.

The excess boundary conditions~(\ref{eq:excess-boundary-condition-0}),~(\ref{eq:excess-boundary-condition})
yield the set of equations
\begin{eqnarray}
\frac{ {\rm i} \kappa\bulk{m} }{ n\bulk{m} } t_p - k_z (1 - r_p) 
& = & -\varepsilon_0 \omega\surresist{} \frac{k^2}{2} \left[ n\bulk{m} t_p + 1 + r_p \right] 
\, ,\\
n\bulk{m} t_p  -  (1 +r_p) 
& = & - \frac{\surconduct}{2\omega\varepsilon_0} 
\left[ \frac{{\rm i} \kappa\bulk{m}}{n\bulk{m}} t_p + k_z (1-r_p) \right] 
\, .
\end{eqnarray}
which is solved for the reflection amplitude 
\begin{equation}
r_p = \frac{\left( \varepsilon\bulk{m} k_z - {\rm i}  \kappa\bulk{m} \right) \left( 1 + \frac{1}{4} \surconduct \surresist k^2 \right) + {\rm i} \frac{\surconduct}{\varepsilon_0 \omega} k_z \kappa\bulk{m} - \varepsilon_0 \varepsilon\bulk{m} \omega \surresist k^2}{\left( \varepsilon\bulk{m} k_z + {\rm i}  \kappa\bulk{m} \right) \left( 1 + \frac{1}{4} \surconduct\surresist k^2 \right) + {\rm i}\frac{\surconduct}{\varepsilon_0 \omega} k_z  \kappa\bulk{m} + \varepsilon_0 \varepsilon\bulk{m} \omega \surresist k^2} \label{eq:reflection_TM} \, .
\end{equation}
Compared with the Fresnel formula~(\ref{eq:Fresnel_TE}), 
this expression features three additional terms.
The familiar Fresnel terms are paired with a mixed term containing surface conductivity and resistivity. Both contribute additional corrections as well. Interestingly, the terms including the resistivity are paired with the parallel component $k$ of the wave vector, suggesting that
this may be a minor correction in the long-wavelength limit ($1/k$ much longer than the
selvedge thickness).

The same calculations can be done for the s-polarisation and lead to
\begin{equation}
r_s = \frac{k_z - {\rm i}  \kappa\bulk{m} - \omega\mu_0 \surconduct(\omega)}{k_z + {\rm i}  \kappa\bulk{m} + \omega\mu_0 \surconduct(\omega)}
\label{eq:reflection_TE}
\end{equation}
that depends only on the surface conductivity, since there is no normal electric field
component in this case.

\begin{figure}[bth]
\centerline{%
\includegraphics*[width=0.6\textwidth]{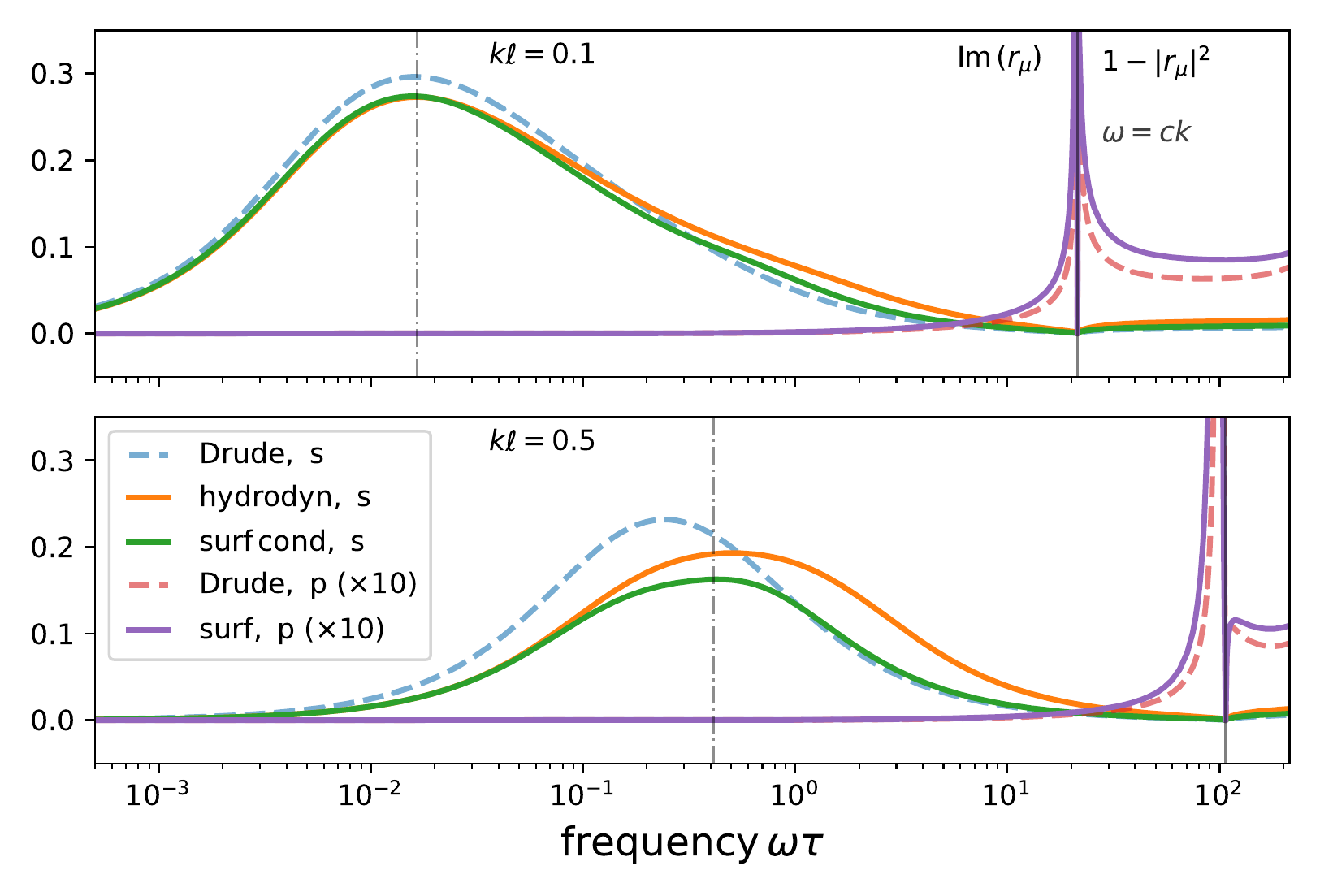}
}
\caption[]{Reflection amplitudes computed with three different models for two values of $k$ (top
and bottom). Both s- and p-polarizations are shown. The light line (vertical line at $\omega = c k$) divides each panel into evanescent (left) and propagating (right) waves. On the left of it, the imaginary part of the reflection amplitude is shown, on the right the absorption. 
The low-frequency maximum in the s-polarization appears at 
$\omega \approx 4.11 (k\lambdabar_p)^2 / \tau$ 
(dash-dotted lines). 
The p-polarization (scaled up by factor 10) gives negligible contributions, except for 
a surface plasmon-like pole near the light line. 
Material parameters for gold (see main text).}
\label{fig:overview-rs-rp}
\end{figure}

For the surface quantities $\surconduct$ and $\surresist$, we propose a Drude-like model
with a relaxation time $\tau_s$ and a length $\ell_0$ that captures 
the thickness of the selvedge region:
\begin{equation}
\surconduct(\omega) = \frac{\surfacelength\sigma_0}{1- {\rm i}\omega \tau_s} \, , 
\qquad
\surresist(\omega) = \surfacelength \frac{1- {\rm i}\omega \tau_s}{\sigma_0}\, .
\label{eq:simple-model-surface-conductivity}
\end{equation}
By comparing to the no-slip viscous model, we estimate $\ell_0 \sim -\ell$, the negative
sign translating the ``missing'' current due to the boundary condition
(Fig.\,\ref{fig:current-profile}).
The behaviour of the reflection coefficients in the far infrared is illustrated in 
Fig.\,\ref{fig:overview-rs-rp}: we note a reduction compared to the local approximation for $r_s$.
This may be attributed to a better impedance matching when the jump of the 
current density at the surface is reduced. A good agreement with the hydrodynamic model 
is found in the long-wavelength limit for
the parameter combination $\ell_0 \approx -0.36\,\ell$ and $\tau_s \approx 2 \tau$.
At larger $k$-vectors, some discrepancies occur. The p-polarization does not contribute
significantly in the evanescent wave sector (away from the light line $\omega = c k$).

\begin{figure}[bth]
\centerline{%
\includegraphics*[width=0.7\textwidth]{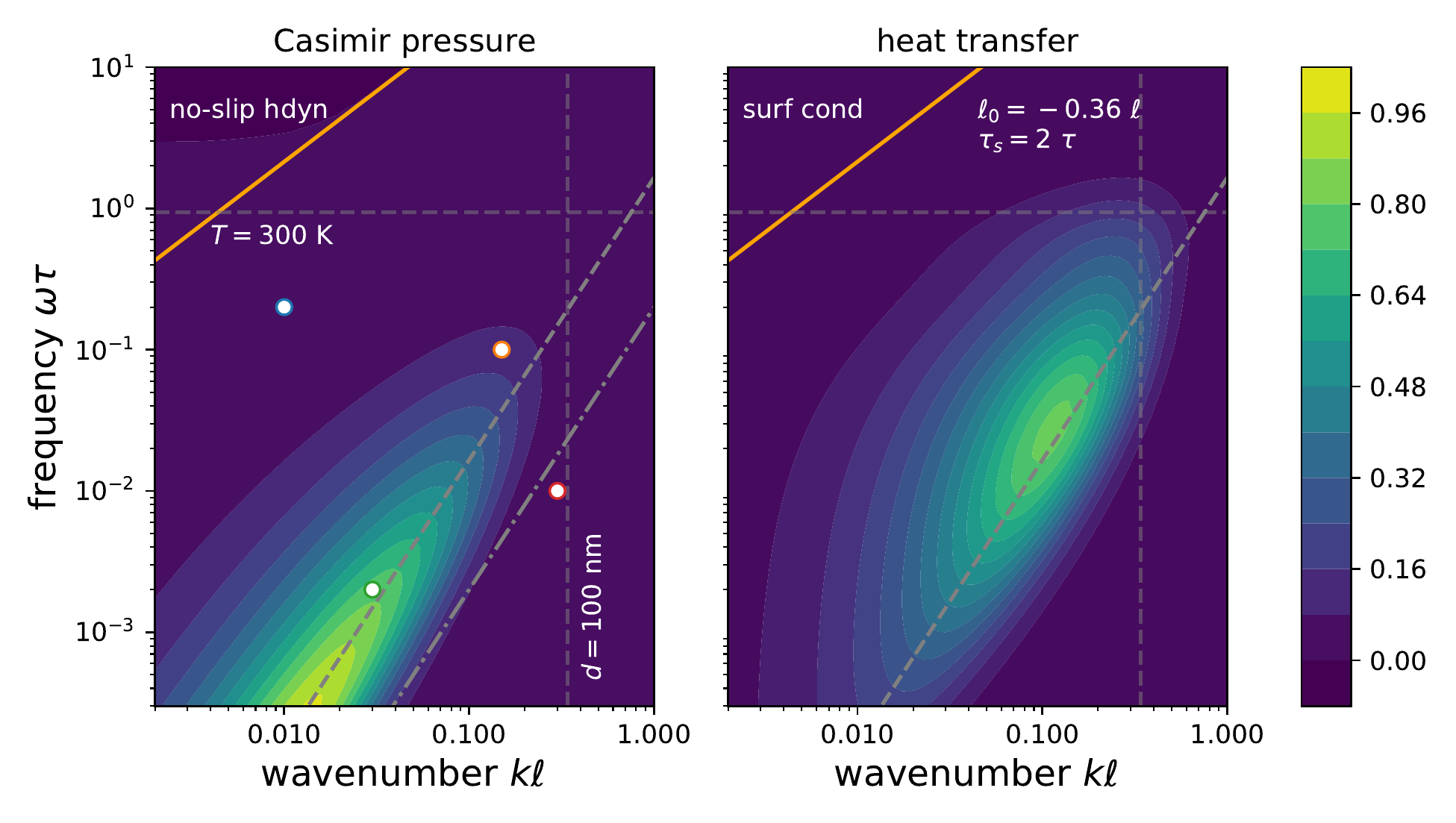}
}
\caption[]{Integrand of the Casimir pressure (left) and heat transport 
(right) in the $k\omega$-plane.
In the left (right) panel, the hydrodynamic (surface conductivity) model is used
to compute the reflection coefficients. Only s-polarized modes are taken into
account. Propagating modes appear above the orange line.
The gray dashed lines mark relevant parameters: temperature ($\hbar\omega = k_B T$), 
distance ($k = 1/d$),
maximum contribution of magnetic diffusion (diagonal dashed) at
$\omega \approx 4.11 (k \lambdabar_p)^2/\tau$, 
diffusive behaviour due to the low-frequency kinematic viscosity
$\omega = 0.2 (k v_{\rm F})^2 \tau$ (dot-dashed).
The data are scaled to the maximum values computed in the local (Drude-Fresnel)
approximation. The white dots in the left mark the values chosen
in Fig.\,\ref{fig:current-profile}.
}
\label{fig:map_both}
\end{figure}

\section{Discussion of results}
\label{s:discussion}

In Fig.\,\ref{fig:map_both}, we plot in the $k \omega$-plane a ``spectral representation'' of the Casimir pressure (left) and the radiative heat transfer (right). Only the thermal contribution of the s-polarization is shown. The data are normalized to the maximum value (in the chosen domain) of the local (Drude-Fresnel) approximation. One notes for both quantities an upper limit $\omega < k_BT / \hbar \sim 1/\tau$, as expected from the Bose-Einstein distribution. The heat transfer data are shifted upwards in frequency due to the additional factor $\omega$ under the integrals~(\ref{eq:heat-transfer-evan}, \ref{eq:heat-transfer-prop}). The maximum in the $k\omega$-plane is set by the magnetic diffusion constant $\lambdabar_p^2/\tau$ (dashed gray lines). A reduction of the pressure appears notably in the range set by the kinematic viscosity $\eta(0) = \frac15 \ell^2/\tau$ (dash-dotted line), when compared to the local approximation. For both quantities, the hydrodynamic and the surface conductivity models give qualitatively the same distributions with the parameters $\ell_0$, $\tau_s$ [Eq.\,(\ref{eq:simple-model-surface-conductivity})] mentioned before [see Fig.\,\ref{fig:map_both}\,(right)].

To summarize, in this paper we have extended the classic Fresnel formulas for the reflection of electromagnetic waves by a metal surface. Two methods have been used: a hydrodynamic description that captures the spatial dispersion of the metal's dielectric function, and a boundary layer technique introducing surface layers of charges and currents. Both methods build on the assumption that the electric current density right at the surface vanishes. This boundary condition corresponds to the behaviour of a viscous fluid, and mirrors the impact of surface roughness on the few-nm scale. The viscosity of conduction electrons was derived from a modification of the well-known Lindhard dielectric functions, taking into account collisions with impurities, but neglecting exchange-correlation effects \citep{Conti_1999}. Note that the no-slip condition is needed to solve the hydrodynamic Navier-Stokes equation that involves higher derivatives of the electronic velocity field. It corrects the spatial profile of the current density right below the surface, and the ``missing current'' is mapped in the boundary layer technique onto a tangential surface current sheet. 
An interesting consequence is a modification of the s-polarized reflection coefficient such that 
repulsive contributions to the Casimir pressure between metallic plates are reduced in both models. This brings theoretical predictions closer to the observed values, possibly pointing towards a physically motivated solution of the so-called ``plasma vs.\ Drude'' controversy.

Among similar attempts to modify the reflection amplitudes by taking spatial dispersion into
account, we mention \citet{Reiche_2020a} and \citet{Klimchitskaya_2020b}. For both, 
the starting point are the surface impedances of a metallic half-space based on a specular 
reflection boundary condition \citep{garcia1979introduction, Ford_1984} where the longitudinal
and transverse dielectric functions appear. 
\citet{Reiche_2020a} use the nonlocal
Lindhard theory corrected for collisions as in Appendix~\ref{a:Lindhard-Conti}, 
but also focus on the impact of Landau damping and the low-temperature behaviour of 
Casimir interactions.
These authors have stressed as well that surface roughness on the scale of the mean free path 
may conflict with the specular reflection assumption.
\citet{Klimchitskaya_2020b} invoke the breaking of translational symmetry due to the 
surface to introduce an anisotropic ${\bf q}$-dependence into the dielectric function.
This correction uses the same small parameter as our hydrodynamic approximation, but is
otherwise quite different in form.
The analysis presented here complements both approaches. 
The Navier-Stokes model allows to resolve spatial nonlocality on scales larger than
the mean free path and predicts nontrivial variations in the sub-surface current when
the no-slip boundary condition is applied.
The excess field (boundary layer) technique collects some of the nonlocality 
into a modified surface response, while allowing for a simpler, local description of the 
bulk. This illustrates that the term `surface' depends on the choice of length scales 
implicit in the formulation of fields and boundary conditions.

We conclude with a few remarks. 
--
The hydrodynamic model has been used in metals long before, but the focus was almost exclusively on the longitudinal response (charge density waves). The corresponding speed of sound $\beta(\omega)$ was derived by \citet{Halevi_1995}. Our analysis links it to complex visco-elastic moduli \citep{Conti_1999} and provides an additional interpretation. The dispersion of $\beta(\omega)$ is actually due to the complex shear modulus of the collisional electron gas in the Navier-Stokes equation, while the bulk viscosity vanishes completely in the hydrodynamic approximation (Stokes hypothesis).

Finally, we expect that the boundary conditions considered here will also modify the surface plasmon dispersion relation (that appears as a peak in $\mathop{\rm Im} r_p$ in Fig.\,\ref{fig:overview-rs-rp}). This is probably irrelevant to radiative heat transfer because it appears in the frequency range $\omega \sim \OmegaP$ where thermal occupation is negligible. The plasmon dispersion has been studied since a long time \citep{garcia1979introduction, Halevi_1995} and depends on the spatial profile of the surface charge on the Thomas-Fermi scale $v_{\rm F} / \OmegaP$. We thus do not expect large modifications since the no-slip condition changes the current density on the much longer scale of the mean free path $v_{\rm F} \tau$, but a quantitative analysis is beyond the scope of this paper.

\vspace{6pt}

\paragraph{Acknowledgements.}
C.H. is indebted to L. P. Pitaevskii for suggesting the dirty limit as a meaningful
simplification of material parameters. 
We thank K. Busch for a careful, constructive reading of the manuscript.
G. W. gratefully acknowledges funding by the German Research Foundation (DFG) in
the framework of the Collaborative Research Center~1375 ``Nonlinear Optics down to Atomic Scales (NOA)''.

\appendix

\section{Derivation of hydrodynamic parameters}
\label{a:Lindhard-Conti}

For the ease of comparison to other work, we first write down the dielectric functions
that follow from the hydrodynamic expressions obtained 
with the Navier-Stokes model~(\ref{eq:Navier-Stokes}) of the electronic liquid:
\begin{eqnarray}
\varepsilon^{\rm hd}_{L}({\bf q}, \omega) &=& 1 - \frac{\OmegaP^2}{\omega(\omega + {\rm i} /\tau) - [{\beta}^2 - {\rm i} \omega (\zeta + \tfrac43\eta) ] q^2}
\label{eq:appendix_longitudinal_hd_epsilon} \\
\varepsilon^{\rm hd}_{T}({\bf q}, \omega) & =& 1 - \frac{\OmegaP^2}{\omega(\omega + {\rm i} /\tau) + {\rm i} \omega {\eta} q^2}
\label{eq:appendix_transverse_hd_epsilon}
\end{eqnarray}
(For simplicity, we put the background dielectric constant $\varepsilon\bulk{b} = 1$ in this 
Appendix.)
We expect the hydrodynamic description to be accurate in the semiclassical and long-wavelength
limits. This requires at least the regime $q \ll k_{\rm F}$, the Fermi momentum. It is also
apparent that the inverses $1 / ( \varepsilon_{L,T}( {\bf q}, \omega) - 1)$ are polynomials
in $q^2$. We shall fix their complex coefficients by a corresponding expansion of the 
dielectric functions of the electron gas. For simplicity, we focus on the degenerate case
(temperature much smaller than the Fermi energy $E_{\rm F} = 5.5\,{\rm eV}$) 
and on the self-consistent field (or random phase)
approximation where the dielectric functions are given by Lindhard theory
\citep{DresselGruenerBook, Lindhard_1954}. 

A technical challenge is to take into account
collisions, since we expect their rate $\sim 1/\tau$ to be comparable to $\omega$.
We apply results from \citet{Kliewer_1969a, Mermin_1970, Conti_1999} who combined the
relaxation-time approximation with the longitudinal and transverse Lindhard functions.
This provides to make contact with the local Drude conductivity as well. 
The same dielectric functions have been used by \citet{Reiche_2020a}.
Let us recall that collisions may involve different scenarios. Our focus is on 
impurity scattering that does not conserve total momentum, while carrier-carrier collisions
do. Electron-phonon scattering is probably some intermediate case due to \emph{Umklapp}
processes \cite{AshcroftMerminBook}. 
These mechanisms lead to distinct temperature dependences of $\tau$. For an overview of
the implications for Casimir-Polder interactions, see \citet{Reiche_2020a, Bordag_2017c}.

We match in the following the hydrodynamic parameters 
to the power series for the inverse susceptibilities
\begin{eqnarray}
\frac{ \OmegaP^2 }{ \varepsilon_{L}({\bf q}, \omega) - 1 }
& \approx &
- \omega \tilde\omega + [{\beta}^2 - {\rm i} \omega (\zeta + \tfrac43\eta) ] \, q^2
+ \ldots
\label{eq:appendix_long_wave_longitudinal_hd_epsilon} 
\\
\frac{ \OmegaP^2 }{ \varepsilon_{T}({\bf q}, \omega) - 1 }
& \approx &
- \omega\tilde{\omega} - {\rm i} \omega \eta \, q^2
+ \ldots
\label{eq:appendix_long_wave_transverse_hd_epsilon}
\end{eqnarray}
where $\tilde{\omega} = \omega + {\rm i}/\tau$. Multiplying these expressions with
${\rm i}\tau/\omega$, we obtain the normalized inverse conductivities 
$\sigma_0 / \sigma_{L,T}({\bf q}, \omega)$ given in
Eqs.\,(\ref{eq:hdyn-longitudinal-sigma}, \ref{eq:hdyn-transverse-sigma}).

\subsection{Expansion of Lindhard functions}
\label{sec:appendix_expansion_Lindhard_functions}

We collect here, for the convenience of the reader, the Lindhard formulas with the dielectric
functions of the homogeneous degenerate electron gas \citep{Lindhard_1954}. 
Lindhard introduced the dimensionless variables
\begin{align}
z = \frac{q}{2k_{\rm F}} \quad \text{and} \quad u = \frac{\omega}{qv_{\rm F}}
\end{align}
The longitudinal dielectric function takes the form
\begin{eqnarray}
\varepsilon^0_{L}({\bf q}, \omega) &=& 
1 + \frac{3 \OmegaP^2}{q^2v^2_{\rm F}} f_{L}(z,u) 
\label{eq:appendix_longitudinal_Lindhard_dielectric_function}
\\
\text{with} \quad f_{L}(z,u) &=& 
\frac{1}{2} + \frac{1-(z+u)^2}{8z}\ln \frac{z+u+1}{z+u-1} 
\nonumber\\
&&\phantom{\frac{1}{2}} {} + \frac{1-(z-u)^2}{8z} \log \frac{z-u+1}{z-u-1}
\label{eq:appendix_longitudinal_Lindhard_f_function}
\end{eqnarray}
and the transverse is
\begin{eqnarray}
\varepsilon^0_{T}({\bf q}, \omega) & = & 1 - \frac{\OmegaP^2}{\omega^2}f_{T}(z,u) 
\label{eq:appendix_transverse_Lindhard_dielectric_function} 
\\
\text{with} \quad f_{T}(z,u) & = & 
\frac{3}{8}\left(z^2+3u^2+1\right) 
- \frac{3 \left[1 - (z + u)^2\right]^2}{32z} 
\log\frac{z+u+1}{z+u-1} 
\nonumber 
\\
&& \phantom{\frac{3}{8}\left(z^2+3u^2+1\right) }
{} - \frac{3 \left[1 - (z - u)^2\right]^2}{32z} 
\log\frac{z-u+1}{z-u-1} 
\label{eq:appendix_transverse_Lindhard_f_function}
\end{eqnarray}
The (natural) logarithms are to be evaluated on their principal branchs, approaching
the real frequency axis from above. This can also be denoted by 
$u = (\omega + {\rm i}0)/q v_{\rm F}$ (hence the superscript $0$ in Eqs.\,(\ref{eq:appendix_longitudinal_Lindhard_dielectric_function}, \ref{eq:appendix_transverse_Lindhard_dielectric_function}). The resulting imaginary parts appear in the domains
$u + z < 1$ and $|u - z| < 1 < u + z$ (and are positive there); they vanish for $|u - z| > 1$.

For the matching with the hydrodynamic expressions~(\ref{eq:appendix_long_wave_longitudinal_hd_epsilon},
\ref{eq:appendix_long_wave_transverse_hd_epsilon}), we perform a double expansion
in the Lindhard variables: small $z$ and large $u$. In this limit, the imaginary parts
do not play a role, and we obtain a regular power series whose first few terms are
\begin{eqnarray}
\frac{ \OmegaP^2 }{ \varepsilon^0_{L}( {\bf q}, \omega ) - 1} & \approx &
- \omega^2
+ \frac{3}{5} v_{\rm F}^2 q^2 
+ \frac{12}{175} \frac{ v_{\rm F}^4 q^4 }{\omega ^2}
+ \frac{ v_{\rm F}^2 q^4 }{ 4 k_{\rm F}^2 }
+ \ldots
\label{eq:appendix_longitudinal_Lindhard_epsilon_expansion_omega_tilde}
\end{eqnarray}
Among the last two terms, the first one dominates if we restrict to frequencies
$\omega \ll v_{\rm F} k_{\rm F} = 2 E_{\rm F}/\hbar$. This is well justified for
$\omega \sim 1/\tau$ and a collisional width $\hbar/\tau \ll E_{\rm F}$.
This result is consistent with Lindhard's Eqs.\,(3.5, 3.10) 
apart from the order $1/u^4$ which has been obtained, however, by 
\citet{Arista_1984}
in the same limit (see Table~1 there). 
\citet{Klimchitskaya_2020b} have also used the small parameter $v_{\rm F} q / \omega$
to add correction terms to the local dielectric function of the Drude model. 
Their correction is, however, of the first order and is anisotropic (the wavevector
$k$ parallel to the metal surface is used in place of $q$).

The corresponding expansion of the transverse dielectric function yields
\begin{eqnarray}
\frac{ \OmegaP^2 }{ \varepsilon^0_{T}( {\bf q}, \omega ) - 1} & \approx &
-\omega ^2 + \frac{1}{5} v_{\rm F}^2 q^2 + \frac{ 8 }{ 175} \frac{ v_{\rm F}^4 q^4}{\omega^2}
+ \ldots
\label{eq:appendix_transverse_Lindhard_epsilon_expansion_omega_tilde}
\end{eqnarray}
Lindhard's Eq.\,(3.19) contains a term $z^2/u^2$ (which vanishes in 
our calculations) and stops before the order $1/u^4$.
We conclude that the small parameter for corrections beyond the hydrodynamic approximation
is $(q v_{\rm F} / \omega)^2 \sim (q \ell)^2$ if we focus on $\omega \sim 1/\tau$. Fortunately
enough, they appear with relatively small numerical coefficients.

\subsection{Including collisions}
\label{sec:appendix_expansion_Mermin_function}

\subsubsection{Longitudinal dielectric function}

Kliewer and Fuchs \citep{Kliewer_1969a}, Mermin \citep{Mermin_1970},
and Conti and Vignale \citep{Conti_1999}
have constructed
the collisional form of $\varepsilon_{L}$ based on the requirement that the
electron gas relaxes to a state defined by a shifted electrochemical potential
$E_{\rm F} + e \phi$ where $\phi$ is computed self-consistently from the induced
charge density. This argument can be carried out for a broad class of dielectric
functions, and for the Lindhard function introduced above, it gives the formula
\begin{equation}
\frac{ \tilde\omega }{ \varepsilon^{\tau}_{L}( {\bf q}, \omega ) - 1 }
= 
\frac{ \omega }{ \varepsilon^{0}_{L}( {\bf q}, \tilde\omega ) - 1 }
+
\frac{ {\rm i}/\tau }{ \varepsilon^{0}_{L}( {\bf q}, 0 ) - 1 }
\label{eq:Mermin-formula}
\end{equation}
We denote by the superscript $\tau$ the collisional form. In the first term on the rhs, 
the electric susceptibility is evaluated at the complex frequency $\tilde\omega$, 
while the second one involves
the static susceptibility that is responsible for the screening of a static charge
density. The latter is evaluated from the Lindhard formula~(\ref{eq:appendix_longitudinal_Lindhard_dielectric_function}) 
by taking the limit $u \to 0$
(approaching zero from the upper half of the complex plane), yielding
\begin{equation}
\lim\limits_{u \to 0} f_{L}(z, u) = \frac{1}{2} + \frac{1-z^2}{4z}\ln\left|\frac{z+1}{z-1}\right|
\label{eq:appendix_zero_u_f_longitudinal}
\end{equation}
Expanding for small $q$, one obtains
\begin{equation}
\lim\limits_{\omega \to 0} 
\frac{ \OmegaP^2 }{ \varepsilon^0_{L}({\bf q}, \omega) - 1 }
= 
\frac{q^2v^2_{\rm F}}{3} \left( 1 + \frac{q^2}{12 k_{\rm F}^2} + \ldots 
\right)
\label{eq:appendix_long_wave_static_longitudinal_Lindhard_diel_fctn}
\end{equation}
Adding the two terms in Eq.\,(\ref{eq:Mermin-formula}), the hydrodynamic power 
series~(\ref{eq:appendix_long_wave_longitudinal_hd_epsilon}) becomes
\begin{eqnarray}
\frac{ \OmegaP^2 }{ \varepsilon^{\tau}_{L}({\bf q}, \omega) - 1 }
& \approx &
- \omega \tilde\omega 
+ \left( \frac{ {\rm i} }{ 3 \tilde\omega\tau } + \frac{3 \omega }{5 \tilde \omega} 
\right)
v_{\rm F}^2 q^2 
+ \frac{12}{175} \frac{ \omega v_{\rm F}^4 q^4 }{\tilde\omega^3}
+ \frac{ {\rm i} }{ \tilde\omega \tau }
\frac{v_{\rm F}^2 q^4}{36 k_{\rm F}^2}
+ \ldots
\label{eq:inverse-longitudinal-susceptibility}
\end{eqnarray}
Note that this treatment of collisions is necessary to match the zero'th order term.
The quadratic term yields the complex combination 
${\beta}^2 - {\rm i} \omega (\zeta + \tfrac43\eta)$ spelled out in Eq.\,(\ref{eq:Halevi-beta}).
Its low-frequency limit $\tfrac13 v_{\rm F}^2$ arises from the static 
susceptibility in Eq.\,(\ref{eq:Mermin-formula}).
The second term of order $q^4$ is again negligible compared to the one before.

\citet{Halevi_1995} found his formula for the speed of longitudinal sound waves by
a similar hydrodynamic argument. The present formalism provides a visco-elastic view 
with a splitting into elastic moduli and viscosities (the latter appear as the imaginary
part of Halevi's $\beta^2$). The connection between Navier-Stokes hydrodynamics and
elasticity theory was also made by \citet{Conti_1999}.

\subsubsection{Transverse dielectric function}
\label{a:transverse-epsilon-from-Lindhard}

To include collisions into the self-consistent field approximation, similar
considerations are applied by \citet{Conti_1999}.
If we assume that the total carrier momentum is not 
conserved (as it happens for impurity scattering), then the resulting
susceptibility takes a form slightly simpler than~(\ref{eq:Mermin-formula})
\begin{equation}
\varepsilon^{\tau}_{T}( {\bf q}, \omega ) - 1 = 
\frac{\tilde{\omega}}{\omega} \left[ \varepsilon^0_{T}( {\bf q}, \tilde{\omega}) - 1 \right]
\label{eq:Kliewer-Conti-formula}
\end{equation}
The power series for the comparison to the hydrodynamic form~(\ref{eq:appendix_long_wave_transverse_hd_epsilon}) is thus
\begin{equation}
\frac{ \OmegaP^2 }{ \varepsilon^0_{T}( {\bf q}, \omega ) - 1} =
-\omega \tilde\omega + \frac{\omega}{5 \tilde\omega} v_{\rm F}^2 q^2 
+ \frac{ 8 }{ 175} \frac{\omega v_{\rm F}^4 q^4}{ \tilde\omega^3}
+ \ldots
\label{eq:inverse-transverse-susceptibility}
\end{equation}
and its quadratic term yields the complex shear viscosity $\eta(\omega)$ given in 
Eq.\,(\ref{eq:Lindhard-Conti-eta}).

It is interesting to subtract the contribution of the shear viscosity $\eta$ 
from $\beta^2 - {\rm i} \omega( \zeta + \tfrac43 \eta)$. 
It turns out that only the static term survives
\begin{equation}
\beta^2 - {\rm i} \omega \zeta = 
\left( \frac{ {\rm i} }{ 3 \tau } + \frac{3 \omega }{5} 
- \frac{4 \omega}{15}
\right) \frac{ v_{\rm F}^2 }{ \tilde\omega }
=
\frac{ v_{\rm F}^2 }{ 3 }
\label{eq:complex-bulk-modulus}
\end{equation}
The bulk modulus (as expressed by the real part $\beta^2$) is thus determined by the
static density response, as expected for a compressibility. Its frequency dependence
is negligible, consistent with the remark of \citet{Conti_1999} that the relevant
frequency scale is the much larger Fermi frequency $E_{\rm F}/\hbar$.
The bulk viscosity $\zeta$, however,
vanishes: collisions do not contribute any losses when compressing the electron gas.
This has been observed also for other models of the dielectric response by
\citet{Conti_1999}.
In particular, they went 
beyond the random phase approximation and included exchange correlation ({\sc xc}) effects 
by dynamic local field factors. Apart from the self-consistent field, the Coulomb interaction between electrons is neglected when using Lindhard's transverse dielectric function. A residual footprint of {\sc xc} effects may be encoded in the electronic lifetime $\tau$, however, as soon as the Lindhard functions are generalized to a collisional model.

\subsubsection{De Andrés \& al}
\label{a:deAndres-et-al}

\citet{deAndres_1986} have argued that the construction of \citet{Kliewer_1969a} for
the collisional transverse dielectric function was in error because 
Eq.\,(\ref{eq:Kliewer-Conti-formula}) could not reproduce, in the static limit, the
weak diamagnetism of the electron gas. 
Their reasoning is based on the relation
\begin{equation}
1 - \frac{1}{{\mu}({\bf q}, \omega)} = \frac{\omega^2}{c^2q^2} \left[ \varepsilon_{L}({\bf q}, \omega) - \varepsilon_{T}({\bf q}, \omega)\right]
\label{eq:appendix_relation_magn_permi_diel_fctns}
\end{equation}
between the relative permeability $\mu$ and the dielectric functions. They apply
Eq.\,(\ref{eq:Mermin-formula}) 
with the substitution $\varepsilon \mapsto {\mu}$ to derive a collisional permeability 
$\mu^{\tau}$. The strict analogy between electric and magnetic response relies on the 
assumption that magnetic charges be conserved. It turns out that 
$\mu^{\tau}( {\bf q}, \omega ) - 1$ is for frequencies $\omega \sim 1/ \tau$ 
close to its static value 
$ - \OmegaP^2 / (4 k^2_{\rm F} c^2 )$ much smaller than unity. (The spin
contribution to the magnetic response would only give small corrections 
\citep{Lindhard_1954, deAndres_1986}.)
Relation~(\ref{eq:appendix_relation_magn_permi_diel_fctns}) then predicts
that $\varepsilon_{T,L}^{\tau}$ are practically the same.
We obtain a complex shear viscosity
\begin{equation}
\text{de Andr\'es \& al.:}\quad
\eta(\omega)  = \frac{{\rm i} v^2_{\rm F}}{\omega}\frac{1/3 - 3 {\rm i} \omega\tau/ 5}{1 - {\rm i} \omega\tau} 
+ \frac{\tilde{\omega}^3\tau}{4\omega k^2_{\rm F}}
\label{eq:result-eta-deAndres}
\end{equation}
where the last term is negligible for $\hbar\omega, \hbar/\tau \ll E_{\rm F}$. This is plotted
in dashed gray in Fig.\,\ref{fig:shear-viscosity}.
Note that
the $1/\omega$ pole would yield a finite velocity for acoustic shear waves, quite unexpected
for a liquid. Also the bulk viscosity $\zeta$ [see Eq.\,(\ref{eq:complex-bulk-modulus})] 
would be nonzero in disagreement with the
general observations of \citet{Conti_1999}. We thus believe that the close analogy between
electric and magnetic responses put forward by \citet{deAndres_1986} (magnetic charge
conservation) is not warranted, at least in the low-frequency region.


\end{document}